\documentclass[aps,prl,twocolumn,superscriptaddress,preprintnumbers]{revtex4}

\usepackage{epsf,epsfig,amsmath,amssymb,amsfonts}
\usepackage{color}
\usepackage{slashed}
\usepackage{hyperref}

\newsavebox{\ns}
\newsavebox{\dbrane}

\def\be{\begin{equation}}
\def\ee{\end{equation}}
\def\bea{\begin{eqnarray}}
\def\eea{\end{eqnarray}}

\def\Dslash{\,\,{\raise.15ex\hbox{/}\mkern-12mu D}}
\def\Dbarslash{\,\,{\raise.15ex\hbox{/}\mkern-12mu {\bar D}}}
\def\delslash{\,\,{\raise.15ex\hbox{/}\mkern-9mu \partial}}
\def\delbarslash{\,\,{\raise.15ex\hbox{/}\mkern-9mu {\bar\partial}}}
\def\pslash{\,\,{\raise.15ex\hbox{/}\mkern-9mu p}}
\def\calDslash{\,\,{\raise.15ex\hbox{/}\mkern-12mu {\cal D}}}

\newcommand\diff{\mbox{d}}

\newcommand{\dd}{\diff}

\allowdisplaybreaks

\begin{document}
\title{Testing the Swampland: $H_0$ tension}

\author{Eoin \'O Colg\'ain}
\affiliation{Asia Pacific Center for Theoretical Physics, Postech, Pohang 37673, Korea \\ Department of Physics, Postech, Pohang 37673, Korea}
\author{Hossein Yavartanoo}
\affiliation{CAS Key Laboratory of Theoretical Physics, Institute of Theoretical Physics, \\ Chinese Academy of Sciences, Beijing 100190, China}

\begin{abstract}

\noindent 
The de Sitter Swampland conjecture compels us to consider dark energy models where $\lambda(\phi) \equiv |\nabla_{\phi} V|/V$ is bounded below by a positive constant. Moreover, it has been argued for Quintessence models that the constant $\lambda$ scenario is the least constrained. Here we demonstrate that increasing $\lambda$ only exacerbates existing tension in the Hubble constant $H_0$. The identification of dark energy models that both evade observational bounds and alleviate $H_0$ tension constitutes a robust test for the Swampland program. 

\end{abstract}

\maketitle

\setcounter{equation}{0}

\section{Introduction}
The de Sitter Swampland conjecture \cite{Obied:2018sgi} claims that de Sitter vacua belong to the Swampland \cite{Vafa:2005ui, Brennan:2017rbf} of low-energy effective theories coupled to gravity. More concretely, it has been proposed that the potential for scalar fields must satisfy the universal bound, 
\be
\label{bound}
|\nabla_{\phi} V| \geq \frac{c}{M_{\textrm{pl}}} \cdot V, 
\ee
where $M_{\textrm{pl}}$ is the Planck mass and $c$ denotes a constant of order 1. Despite being controversial - it questions $\Lambda$CDM - the conjecture can be motivated from the distance conjecture \cite{Ooguri:2006in} and the Bousso covariant entropy bound \cite{Bousso:1999xy}, which places it on firmer footing \cite{Ooguri:2018wrx} (see also \cite{Garg:2018reu, Andriot:2018mav, Brahma:2019mdd}).  

The cosmological implications of the conjecture were initially studied in \cite{Agrawal:2018own}, where it was suggested that Quintessence models \cite{Copeland:2006wr} with an exponential potential $V = V_0 e^{-\lambda \phi}$, where $V_0$ and $\lambda$ are constants, constitute valid dark energy models,  valid in the sense that they satisfy the bound (\ref{bound}) and are the least constrained by data. The constraints from data were subsequently tightened in \cite{Heisenberg:2018yae, Akrami:2018ylq, Raveri:2018ddi} (also \cite{Agrawal:2018rcg, Elizalde:2018dvw}). Throughout, the potential elephant in the room has been ``$H_0$ tension", a discrepancy between a local determination of the Hubble constant due to Riess et al. \cite{Riess:2016jrr, Riess:2018byc, Riess:2019cxk} and Planck CMB analysis based on $\Lambda$CDM \cite{Aghanim:2018eyx}. Indeed, in the lifetime of the de Sitter Swampland conjecture, we have witnessed the statistical importance of the difference slowly tick up from $3.8 \, \sigma$ \cite{Riess:2018byc} to $4.4 \, \sigma$ \cite{Riess:2019cxk}. At some point this becomes difficult to ignore. 

In the immediate aftermath of the conjecture \cite{Obied:2018sgi}, we suggested it was natural if $H_0$ tension and the Swampland were connected \cite{Colgain:2018wgk}. More precisely, in order to reconcile the Riess result with other cosmological determinations of Hubble at low redshift $z < 2$, we floated the idea of a turning point in $H(z)$. Explaining the tension remains an open problem, but in this letter we take aim at the models proposed in \cite{Agrawal:2018own} and take them to their logical conclusion by asking, do they reduce $H_0$ tension? The answer seems no. Note, there are various recent studies of Quintessence in the literature and the tension cannot be fully relieved \cite{DiValentino:2017zyq, Durrive:2018quo, Yang:2018xah, Tosone:2018qei}, so this conclusion is not overly surprising. However, see \cite{Agrawal:2019dlm} for recent progress reconciling the de Sitter Swampland with $H_0$ tension.

Our analysis here is simple and to the point. It emphasises the need to consider $H_0$ tension as a further litmus test for good dark energy models satisfying (\ref{bound}). Otherwise, the Swampland may be in jeopardy. The simplicity of our analysis comes from the fact that we can piggy-back on \cite{Agrawal:2018own} and integrate a single additional first-order differential equation to identify the Hubble parameter as a function of redshift $H(z)$. 

Of course, integrating a first-order ODE requires a boundary condition. Here, we assume $H(z = 0) = H_0$, where $H_0$ is a constant, and use a well-known compilation of cosmological measurements of $H(z)$ at low redshift \cite{Farooq:2016zwm}, as well as priors from Planck $H_0 = 67.4 \pm 0.5 \textrm{ km s}^{-1} \textrm{ Mpc}^{-1}$ \cite{Aghanim:2018eyx} and Riess et al. $H_0 = 74.03 \pm 1.42 \textrm{ km s}^{-1} \textrm{ Mpc}^{-1}$ \cite{Riess:2019cxk}, to determine $H_0$ through a best-fit to the data. For simplicity, we follow the analysis of \cite{Agrawal:2018own} (also \cite{Heisenberg:2018yae}) and assume the dark energy density today is 70\%. Doing so, we will see that as $\lambda$ increases from $\lambda = 0$ to $\lambda =1$, the tension in the Hubble constant increases from $4.3 \, \sigma$ to $4.7 \, \sigma$ with a Planck prior and increases from $2.8 \, \sigma$ to $3.9 \, \sigma$ with a Riess et al. prior on $H_0$. 

Regardless of the prior, throughout this analysis we quote tension between Quintessence and the higher Riess et al. value \cite{Riess:2019cxk}, since the Hubble constant is a parameter one should measure locally that is in principle independent of any assumed cosmology. An obvious caveat with this approach is that we may live in a local cosmic void \cite{Shanks:2018rka} and $H_0$ determined this way may not be representative. This concern aside, in light of recent independent H$0$LiCOW results that also favour a higher $H_0$ \cite{Wong:2019kwg}, there is now less reason to doubt the local determination.

\noindent
\section{Quintessence}
Here we quickly review Quintessence following \cite{Tsujikawa:2013fta}. Consider the scalar-gravity action 
\be
S = \int \dd^4 x \sqrt{-g} \left[ \frac{1}{2} M_{\textrm{pl}}^2 R - \frac{1}{2} \partial_{\mu} \phi \partial^{\mu} \phi - V(\phi) \right] + S_m, \nonumber
\ee
where $S_m$ denotes the matter sector. Consider also the Ricci-flat Friedmann-Lema\^itre-Robertson-Walker (FLRW) background with scale factor $a(t)$. Assuming non-relativistic matter, the equation of state $w_m$ for the matter sector becomes $w_m = 0$, so we can regard matter as pressureless $p_m = 0$. With this restriction the equations of motion of the above action can be succinctly recast in terms of the following dynamical system, 
\bea
\label{eq1} x' &=& - 3 x + \frac{\sqrt{6}}{2} \lambda y^2 + \frac{3}{2} x ( 1 + x^2 - y^2), \\
\label{eq2} y' &=&  - \frac{\sqrt{6}}{2} \lambda x y + \frac{3}{2} y (1 + x^2 - y^2), \\
\label{eq3} \frac{H'}{H} &=& - \frac{3}{2} (1 + x^2 - y^2), 
\eea
where we have defined the variables: 
\be
x \equiv \frac{\dot{\phi}}{\sqrt{6} M_{\textrm{pl}} H}, \quad y \equiv \frac{\sqrt{V(\phi)}}{\sqrt{3} M_{\textrm{pl}} H}. 
\ee
Note that dots denote the usual time derivatives and primes denote derivatives with respect to $N \equiv \ln a$. In addition, we have defined $\lambda \equiv - M_{\textrm{pl}} \nabla_{\phi} V /V$. In terms of these new variables the scalar density $\Omega_{\phi} \equiv \rho_{\phi}/(3 M_{\textrm{pl}}^2 H^2)$ describing dark energy is 
\be
\Omega_{\phi} =x^2 + y^2, 
\ee
which sums to unity with the matter density $\Omega_m \equiv \rho_{m}/(3 M_{\textrm{pl}}^2 H^2)$, i. e. $\Omega_m + \Omega_{\phi} = 1$. 

As explained in \cite{Tsujikawa:2013fta}, when $\lambda$ is constant, the system has four fixed points where $x' = y' = 0$. When $w_{m} = 0$ two fixed points are suitable for modeling the matter-dominated regime $\Omega_{\phi} \ll 1$, namely $(x,y) = (0,0)$ and $(x, y) = (\sqrt{3/2}/\lambda, \sqrt{3/2}/\lambda$), but for the latter one requires the additional condition $\lambda \gg 1$. Cosmic acceleration can be realised at a third fixed point $(x, y) = (\lambda/\sqrt{6}, \sqrt{1- \lambda^2/6})$, but only when $\lambda^2 < 2$. For this reason, the interpolating flows from the vicinity of unstable saddle point $(x, y) = (0,0)$, where $\Omega_{\phi} = 0$, to the stable fixed point $(x, y) = (\lambda/\sqrt{6}, \sqrt{1- \lambda^2/6})$, with $\Omega_{\phi} =1$, provide us with one of the simplest models to describe a transition from a matter-dominated regime to one of cosmic acceleration. 

Following \cite{Agrawal:2018own, Heisenberg:2018yae}, we focus on this trajectory. To be more precise, we integrate (\ref{eq1}) and (\ref{eq2}) in $N$ from a point close to $(x, y) = (0,0)$ to a point $N_*$ where $\Omega_{\phi} = 0.7$, which corresponds to a dark energy density of 70\% today. We then employ the redefinition $N - N_* = -\ln(1+z)$, so that $N = N_*$ corresponds to redshift $z = 0$ today. To further identify $H(z)$, we change variables from $N$ to $z$ and integrate (\ref{eq3}) in $z$ subject to the condition that $H(0) = H_0$, where $H_0$ is a constant we will later determine through fits to the cosmological data. For presentation purposes, essentially to mirror the analysis of \cite{Agrawal:2018own, Heisenberg:2018yae}, we restrict our data fitting to $z < 1$, but one can check that extending this out to $z = 2.36$ does not change conclusions. 

Before proceeding to data fitting in the next section, let us attempt to anticipate the end result. First, let us change variables in equation (\ref{eq3}) to redshift $z$:
\be
\frac{(1+z) H'(z)}{H} = \frac{3}{2} ( 1 + x^2 - y^2).  
\ee
What is important here is the slope of the Hubble parameter $H(z)$, i. e. $H'(z)$. If $H'(z)$ increases with $\lambda$, then we can expect the Farooq et al. data \cite{Farooq:2016zwm} to pull $H_0$ downwards to compensate, and alternatively, if $H'(z)$ decreases, then the value of $H_0$ will increase. By fixing a nominal value for $H_0 = 70 \textrm{ km s}^{-1} \textrm{ Mpc}^{-1}$, it is easy to see from Figure 1 that the slope increases with $\lambda$. For this reason, it is a foregone conclusion that $H_0$ will decrease with increasing $\lambda$. In the next section, we will study this effect using a concrete dataset and confirm that for the data we consider $H_0$ tension increases.

\begin{figure}[h]
\begin{center} 
    \includegraphics[scale=0.8]{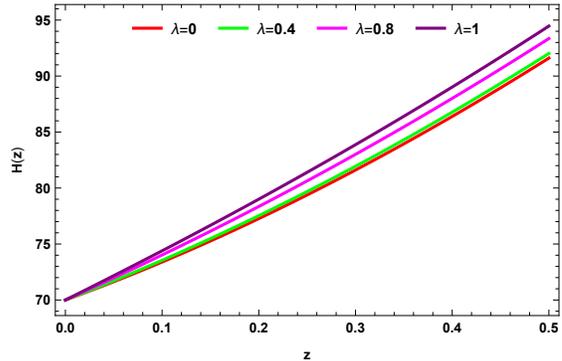}
   \caption{Integrating equation (\ref{eq3}) subject to the nominal boundary condition $H_0 = 70 \textrm{ km s}^{-1} \textrm{ Mpc}^{-1}$ leads to an increasing slope with increasing $\lambda$.}
 \end{center}
\end{figure}

\noindent 
\section{Cosmological data} 
The data we use in this study comprises cosmological measurements of the Hubble parameter $H(z)$ compiled by Farooq et al. \cite{Farooq:2016zwm} based on cosmic chronometric and  baryon acoustic oscillations (BAO) techniques \cite{Simon:2004tf, Stern:2009er, Moresco:2012jh, Blake:2012pj, Zhang:2012mp, Font-Ribera:2013wce, Delubac:2014aqe, Moresco:2015cya, Moresco:2016mzx, Alam:2016hwk}. To this data we will add the Planck prior $H_0 = 67.4 \pm 0.5 \textrm{ km s}^{-1} \textrm{ Mpc}^{-1}$ \cite{Aghanim:2018eyx} and the Riess et al. prior $H_0 = 74.03 \pm 1.42 \textrm{ km s}^{-1} \textrm{ Mpc}^{-1}$  \cite{Riess:2019cxk} in turn and document the result. Even in the absence of a $H_0$ prior the remaining data has a preference for Planck and for this reason we begin our analysis with a Planck prior. This preference should come as little surprise, since the BAO data, which has the smallest errors, assumes a length scale inferred from CMB. 

To get oriented, we recall that $\Lambda$CDM can be described at low redshift by the expression
\be
\label{LCDM}
H(z) = H_0 \sqrt{1 - \Omega_m + \Omega_m (1+z)^3}, 
\ee
where $H_0$ is the Hubble constant and $\Omega_m$ is the matter density today. Recalling that $\Omega_m + \Omega_{\phi} = 1$, we set $\Omega_m = 0.3~ (\Omega_{\phi} = 0.7)$ and perform an error-weighted least squares fit to the $z< 1$ data with Planck prior. This returns the best-fit value
\be
\label{fit}
H_0 = 67.70  \pm 0.42 \textrm{ km s}^{-1} \textrm{ Mpc}^{-1}, 
\ee
where throughout we allow for a $1 \, \sigma$ error. At this point we can comment on the tension with the latest Riess et al. value $H_0 = 74.03 \pm 1.42 \textrm{ km s}^{-1} \textrm{ Mpc}^{-1}$  \cite{Riess:2019cxk}. Using the best-fit value above based on Planck prior and cosmological data \cite{Farooq:2016zwm}, the tension is 
\be
\frac{(74.03 - 67.70)}{\sqrt{(1.42)^2 + (0.42)^2}} \, \sigma \approx 4.3 \, \sigma,  
\ee
in line with the claims of Riess et al.  \cite{Riess:2019cxk}. It is worth noting that the tension has been slightly relaxed from $4.4 \, \sigma$ to $4.3 \, \sigma$ through the introduction of the extra data \cite{Farooq:2016zwm}.

We will use this fit as a consistency check on our numerical solution. This exercise simply ensures that we have integrated (\ref{eq3}) correctly. Setting $\lambda = 0$ and performing a similar error-weighted best-fit of the numerical solution $H(z)$ to the data, we recover the above fit (\ref{fit}). This agreement is illustrated in Figure 2, where we also include the data for comparison.  

\begin{figure}[h]
\begin{center} 
    \includegraphics[scale=0.8]{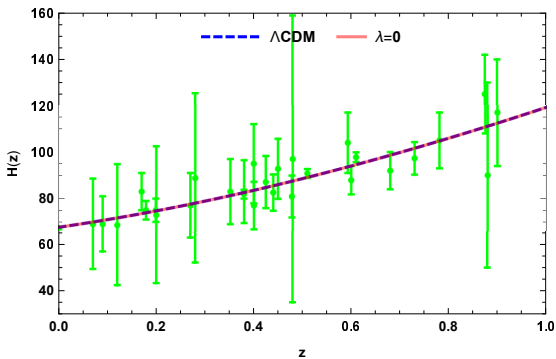}
   \caption{Here we illustrate the consistency between the analytic and numerical solution ($\lambda = 0)$ with a Planck prior. This provides a consistency check on the numerical integration. }
 \end{center}
\end{figure} 

Now that we have confidence in the numerical solution for $\lambda = 0$, namely that it recovers $\Lambda$CDM, we turn our attention to non-zero values of $\lambda$. From the perspective of the de Sitter Swampland conjecture, it is important that $\lambda > 0$, but as $\lambda$ gets larger the constraints from datasets become more stringent \cite{Agrawal:2018own, Heisenberg:2018yae, Akrami:2018ylq, Raveri:2018ddi}, so we do not probe above $\lambda = 1$. 

However, the point we want to make here is different and this brings us to the crux of this short note. Failing some unanticipated systematics, there is growing consensus in the community that the higher Riess et al. local determination of the Hubble constant \cite{Riess:2019cxk} is here to stay. This may provide some hint of a new cosmology beyond $\Lambda$CDM, potentially one without a de Sitter attractor, thus fitting the key premise of the de Sitter Swampland program. That being said, if $H_0$ tension is real, then deviations from $\Lambda$CDM within this framework should not lower the best-fit value of $H_0$ and therefore only increase the tension: they should not make the situation worse.

\begin{table}[h]
\begin{center}
\begin{tabular}{c|cc}
$\lambda$ & $H_0$  & Tension \\
\hline
$0$ & $67.70 \pm 0.42$ &  $4.3\, \sigma$ \\
$0.2$ & $67.67 \pm 0.42$ &  $4.3\, \sigma$ \\
$0.4$ & $67.61 \pm 0.42$ &  $4.3\, \sigma$ \\
$0.6$ & $67.49 \pm 0.41$ &  $4.4\, \sigma$ \\
$0.8$ & $67.32 \pm 0.41$ &  $4.5 \, \sigma$ \\
$1$ & $67.08 \pm 0.41$ &  $4.7 \, \sigma$ \\
\end{tabular}
\caption{Best-fit values of $H_0$ with $\lambda$ subject to a Planck prior.}
\end{center} 
\end{table}

In the above table we record the best-fit values of $H_0$ as $\lambda$ is increased and confirm that $H_0$ is traveling in the wrong direction relative to $\Lambda$CDM. In other words, the tension is increasing. We provide a graphical representation of the same tendency in Figure 3, where we include the current Riess et al. determination $H_0 = 74.03 \pm 1.42 \textrm{ km s}^{-1} \textrm{ Mpc}^{-1}$ \cite{Riess:2019cxk} for reference. As is evident from the plot, as $\lambda$ is increased, the data pulls the best-fit $H_0$ to lower values and thus further away from Riess et al. This behaviour can be contrasted with the two-parameter model presented in \cite{Colgain:2018wgk}, based on \cite{vanPutten:2017bqf}, which favours the higher value against the same data without a Planck prior for $H_0$. 

\begin{table}[h]
\begin{center}
\begin{tabular}{c|cc}
$\lambda$ & $H_0$  & Tension \\
\hline
$0$ & $69.60 \pm 0.66$ &  $2.8 \, \sigma$ \\
$0.2$ & $69.54  \pm 0.66$ &  $2.9 \, \sigma$ \\
$0.4$ & $69.36 \pm 0.66$ &  $3 \, \sigma$ \\
$0.6$ & $69.05 \pm 0.66$ &  $3.2\, \sigma$ \\
$0.8$ & $68.60 \pm 0.65$ &  $3.5 \, \sigma$ \\
$1$ & $67.98 \pm 0.65$ &  $3.9 \, \sigma$ \\
\end{tabular}
\caption{Best-fit values of $H_0$ with $\lambda$ subject to a Riess et al. prior.}
\end{center} 
\end{table}

\begin{figure}[h]
\begin{center} 
    \includegraphics[scale=0.8]{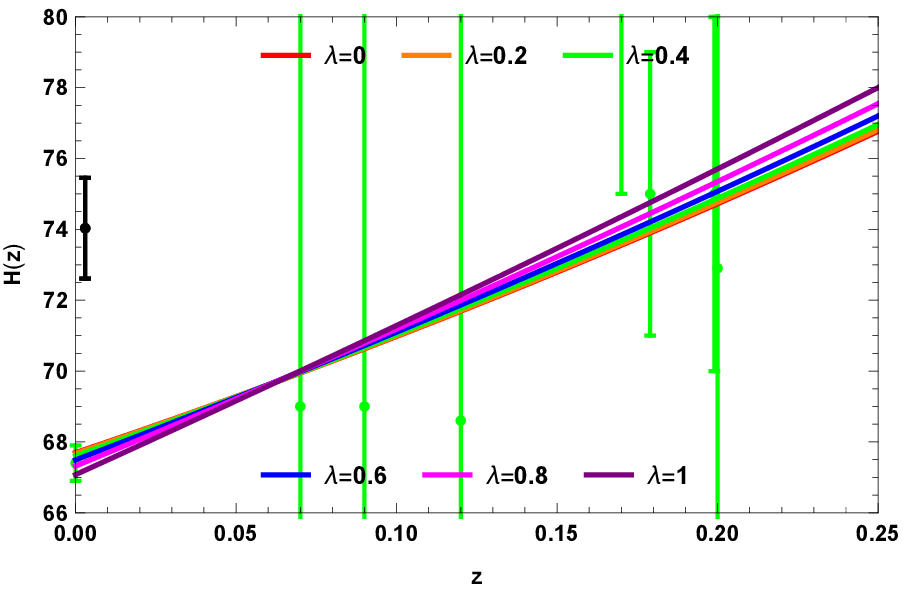}
   \caption{We illustrate the best-fit values of $H_0$ as $\lambda$ is varied subject to a Planck prior for $H_0$. As is evident from the plot, larger values of $\lambda$ lead to lower values of $H_0$. We include the current Riess et al. value $H_0 = 74.03 \pm 1.42 \textrm{ km s}^{-1} \textrm{ Mpc}^{-1}$  \cite{Riess:2019cxk} for comparison.}
 \end{center}
\end{figure}

\begin{figure}[h]
\begin{center} 
    \includegraphics[scale=0.8]{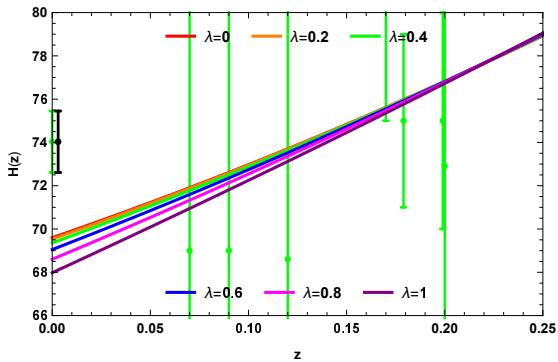}
   \caption{We illustrate the best-fit values of $H_0$ as $\lambda$ is varied subject to a Riess et al. prior for $H_0$. As is evident from the plot, larger values of $\lambda$ lead to lower values of $H_0$. We include the current Riess et al. value $H_0 = 74.03 \pm 1.42 \textrm{ km s}^{-1} \textrm{ Mpc}^{-1}$  \cite{Riess:2019cxk} for comparison.}
 \end{center}
\end{figure}

For completeness, let us now switch the Planck prior for $H_0$ with the Riess et al. result. As we shall see, this changes the numbers, but not the conclusion that increasing $\lambda$ only exacerbates $H_0$ tension. The result of this exercise is illustrated in Table II and Figure 4. Evidently, changing the prior has the expected result of pushing up the best-fit $H_0$ values and leads to a larger error, which ultimately reduces the tension. Nevertheless, the tendency of $H_0$ to decrease with increasing $\lambda$ is the same.

\section{Summary}
To date studies of the de Sitter Swampland conjecture in the context of late-time cosmology have largely overlooked the tension in the Hubble constant. Following up on comments made initially in \cite{Colgain:2018wgk}, the point we wish to drive home here is that the simple Quintessence models considered in \cite{Agrawal:2018own}, and followed up elsewhere, appear to inadvertently make the problem worse. 

To support our case, we made use of cosmological measurements of the Hubble parameter and used this data to infer the Hubble constant. Given the current status of the data, for any monotonically increasing function of the Hubble parameter, essentially a feature that is expected to come from good EFTs where the null energy condition (NEC) is satisfied, one should arrive at similar conclusions, for example \cite{Vagnozzi:2018jhn, Han:2018yrk, Cai:2018ebs, Heckman:2019dsj, Heisenberg:2019qxz, Brahma:2019kch, vandeBruck:2019vzd, Mukhopadhyay:2019cai}. In this sense, we believe that Quintessence is representative, but data may favour more dramatic reconciliations for $H_0$ tension e. g.  \cite{Dutta:2018vmq}. Note, although data may naively point to a violation of the NEC, this can in principle be masking further physics with no violation of the NEC. See \cite{Csaki:2004ha, Csaki:2005vq} for work in the context of dark energy. 


Since $H_0$ tension may be expected to persist in the near future, it is imperative to identify dark energy models that satisfy the Swampland constraints (\ref{bound}), evade observational bounds, but to be prudent, it is best they do not increase $H_0$ tension. This provides a further litmus test for good dark energy models motivated by the Swampland.  See \cite{Agrawal:2019dlm} for recent progress in this direction.

\section*{Acknowledgements}
We thank Robert Brandenberger, Lavinia Heisenberg, Shahin Sheikh-Jabbari  \& Cumrun Vafa for discussion. This work was supported in part by the Korea Ministry of Education, Science and Technology, Gyeongsang-buk Do and Pohang City and  by National Natural Science
Foundation of China, Project 11675244. EOC thanks Mainz Institute for Theoretical Physics (MITP) of the DFG Cluster of Excellence PRISMA$^+$ (Project ID 39083149) for hospitality and support.

\end{document}